\documentclass[jkps,twocolumn,a4paper]{revtex4-1}
\usepackage{graphicx}
\usepackage{amssymb}
\usepackage{amsmath}
\usepackage{bm}
\usepackage{epsfig}
\usepackage{xspace}
\usepackage{paralist}
\usepackage{lipsum}

\newcommand{\U}[1]{\ensuremath{\rm ^{#1}U}\xspace}
\newcommand{\Th}[1]{\ensuremath{\rm ^{#1}Th}\xspace}
\newcommand{\ppt}{\ensuremath{\cdot10^{-12}~g/g}\xspace}
\newcommand{\K}[1]{\ensuremath{\rm ^{#1}K}\xspace}

\newcommand{\KTHU}{K, Th, and U\xspace}
\newcommand{\uph}{ultra-pure \ensuremath{\rm HNO_3}\xspace}
\newcommand{\hnot}{\ensuremath{\rm HNO_3}\xspace}

\newcommand{\cels}{\ensuremath{\rm ^{\circ}C}\xspace}

\begin{document}

\setcounter{page}{0}
\title[]{A convenient approach to $10^{-12}~g/g$ ICP-MS limits for Th and U in Aurubis electrolytic NA-ESN brand copper. }
\author{Douglas \surname{Leonard}}
\email{leonard@uos.ac.kr}
\affiliation{Department of Physics, University of Seoul, Seoul, Korea}

\begin{abstract}
Inductively coupled plasma mass spectroscopy is a powerful technique for measuring trace levels of radioactive contaminants, specifically Th and U, in materials for use in construction of low-background rare-event detectors such as double beta decay and dark matter detectors.  I describe here a technique for measuring Th and U contamination in  copper using direct acid digestion and dilution, without further chemical processing, achieving results comparable to previous work~\cite{grinberg,Leonard08} which utilized more complex chemical pre-concentration techniques.  A convenient research-oriented analysis environment is described as well. Results are presented for measurements of three samples from the production line of electrolytically-purified, LME (London Metal Exchange) grade A, NA-ESN Aurubis copper.  Purified samples showed levels consistent with zero contamination for both elements, while weak but inconclusive indications of contamination were present for the un-purified anode copper.  The best limits achieved are near 1\ppt (95\% CL) for both Th and U measured for copper from the cathode of the purification process.   

\end{abstract}



\maketitle

\section{INTRODUCTION and BACKGROUND}

Many recent advances and endeavors in particle physics, specifically neutrino and dark‐matter physics, involve searches for events with extraordinarily low interaction rates. Event rates are enhanced by increasing detector sizes and/or increasing the source size or strength (detector material being the same as the source in many cases).  The lack of sufficiently specific event signatures, coupled with the need to improve sensitivities, often demands strict controls on many sources of mundane background events which can potentially overwhelm the signals of interest. 

Controlling background rates in next-generation experiments requires meticulous and extensive quantification of all potential background sources, including naturally abundant radioactive isotopes within the detection medium and surrounding materials. Experimental impacts of the quantified background sources are typically estimated by Monte Carlo simulation. The results feed directly into actionable decisions and are a driving constraint in nearly every aspect of the experiments including  materials selection, detector location, design, construction, operation, and cost. 

Much of the focus of background control is on the prevalent and long lived naturally occurring radioactive isotopes of \KTHU.  All three can be problematic, but \Th{232}and \U{238} have significantly higher decay energies than \K{40}, making them a concern in a wider range of experiments and at more outer layers of shielding relative to \K{40}.  Specifically, isotopes used in most neutrinoless double beta decay searches have end-point energies far above the 1460~KeV Q-value of the \K{40} decay.  Several methods are used to detect these radioactive isotopes.  In the case of \K{40}, its abundance can alternatively be inferred by measurement of the non-radioactive isotopes of K, which have natural abundances four orders of magnitude higher, thus drastically reducing sensitivity requirements for some techniques.  

The most direct assay method is clearly to detect the natural radioactivity itself, usually using a germanium counter, but this can require large amounts of materials, typically on the scale of 1~kg, and long counting time, often on the scale of two weeks, where the primary detection apparatus is fully utilized with a single sample. To be relevant sensitivities must, in many cases, compete with those of the next-generation detector being designed.  Neutron activation analysis (NAA) can drastically increase sensitivities by exposing samples to a neutron flux in a nuclear reactor, producing short lived isotopes with far higher activities than those initially present. This process can achieve higher sensitivities, often around 1\ppt for Th and U, with smaller sample masses and less counting time per sample relative to direct counting. NAA has disadvantages though.   Use of the reactor  can be expensive and  either requires analysis facilities at the reactor or requires careful transport and handling of radioactive material in compliance with relevant regulations. NAA is more labor intensive than direct counting, but most importantly, interference signals render direct NAA analysis essentially useless for many types of materials.  

Inductively coupled plasma mass spectroscopy (ICP-MS) detects isotopes by atomic mass, usually after acid digestion of a small sample, typically with masses of a few grams or less. The process is especially suitable for many metals which are digestible in common acids. This compliments NAA well since NAA  interferences are high for most metals such as aluminum, copper, and gold.  Furthermore many of the nuts and bolts, literally,  used in detector construction are most easily and desirably constructed of metal.  ICP-MS analysis requires more sample preparation work than direct counting, and more labor, but occupies the detection apparatus for a much smaller time. This allows high throughput rates of analysis samples.  Unlike direct counting, both ICP-MS and NAA measure concentrations of progenitor isotopes, not the daughters in the decay chains which directly produce the background radiation.  Interpretation of these results thus requires an assumption that the decay chain is in secular equilibrium.   This assumption can be violated if isotopes within the decay chain were separated from each other by chemical or physical processes. 

To set limits near $10^{-12}~g/g$ in Th and U, previous ICP-MS work for the EXO experiment~\cite{Leonard08,grinberg} used complex chemical pre-concentration methods.  By separating and removing the matrix (primary sample material), higher concentrations of contaminants can be introduced into the apparatus for analysis.  These methods  introduce more handling steps, result in contact with more surfaces, and require use of more chemicals. These all complicate preparation and process verification and produce many extra sources for contamination of the samples and process blanks. Ultimately sensitivity can be limited by increased contamination of process blanks rather than by the concentration enhancement. This method was used extensively for rapid screening of many metallic materials used in EXO-200, contributing to production of a detector with one of the lowest background levels ever achieved~\cite{Leonard08,EXO2nu1,EXOpartI,EXO0nu,EXO2nu2}.  There are concerns over secular equilibrium, as mentioned, and also potentially with variation in sensitivity depending on the chemical form of the contaminants. However, in many cases no other technique can reach the desired sensitivities and the practical application of the technique has proven very effective.  Neither of these concerns will be directly addressed in this work. 

I describe here development of an environment and method for measuring 1\ppt limits for Th and U levels in copper, specifically applied to Aurubis LME (London Metal Exchange) grade A, NA-ESN brand copper, using a minimal laboratory environment and a simplified direct-digestion ICP-MS analysis procedure. Electrolytic copper is a common construction material for large detector parts largely because it has been shown repeatedly to be available with high radio-purity, and for its clearly useful combination of mechanical, thermal and electrical properties. Aurubis (formerly Norddeutsche Affinerie AG)copper in particular was qualified for~\cite{Leonard08} and used in~\cite{EXOpartI} the above mentioned EXO-200 detector. The efforts described here are largely inspired by the previous work of Ref.~\cite{Leonard08} and ~\cite{grinberg} but with a focus on achieving similar results with simplified methods.

\section{Laboratory Setup}

A trace analysis ICP-MS lab was prepared for the purpose of economical bench-top research.  The space is defined by a soft-walled cleanroom with nominally uni-directional flow with outside dimensions of approximately  4.9~m $\rm\times$ 3.2~m including a small gowning area.  A unidirectional flow design, with its advantage of efficient air-turnover, is well implemented with a soft-walled cleanroom.  Influx of dirt across the floor from outside the cleanroom can be minimized by good cleanliness practices. A soft-walled cleanroom, when suitable, has some benefits over a hard-walled cleanroom, including simpler and cheaper construction, the ability to quickly reconfigure facilities (electrical, water, gas, etc), simpler access to equipment at all sides of the cleanroom for maintenance etc, and more flexibility in moving equipment in and out of the cleanroom.  These same advantages can violate good cleanliness practices if used carelessly but are useful conveniences for making non-routine changes in a small-scale dynamic research environment.  Two walls of the cleanroom were placed approximately 80~cm away from the concrete laboratory walls to allow access to these cleanroom walls, to allow space for gas bottle storage and usage, and to allow air to flow out and up from the bottoms of these walls. 

The cleanroom frame was constructed using hot-dipped galvanized u-channel steel which can easily be cut and joined with hand tools.  Steel was chosen based on cost, strength, and safety by way of its ability to absorb energy even beyond its failure point.  Standard double U-channel (80~mm $\rm\times$ 40~mm profile, 2.6~mm thick) was selected to provide sufficient strength for the desired spans in abusive use conditions.   Posts were bolted to the floor, and lateral and torsional stability was added by bracing the top of the frame to the two adjacent walls in five locations.  Sidewall curtains were hung with overlapping sheets of anti-static vinyl of 0.5~mm thickness hanging to within about 10~cm from the floor.  Ceiling tiles were cut from 3~mm polyethylene.  The primary work surface is glass, chosen for cost, chemical resistance and ease of cleaning.
Air flow has been provided by three standard fan filter units (FFUs) distributed across the working area of the cleanroom, each with approximately 120~cm x 60~cm filter area and 0.45~m/s rated air flow.  This provides an estimated average flow rate in the working area of about 5.6~m/min providing about 150 air changes per hour, meeting guidelines for ISO 6 (class 1000) cleanrooms~\cite{Cleanrooms}.

Finally the cleanroom houses a PerkinElmer NexION 300-Q ICP-MS machine for trace analysis, as well standard accessories for sample cleaning and handling.  These include an ultra-pure water supply (Millipore Direct-Q 3 UV with 30~L tank), a large chemical resistant fume hood ($\rm 180~cm \times 800~cm$ working area), an analytic balance, and a heated ultrasonic cleaner. 

High flow of exhaust air from equipment can potentially overwhelm the FFU flow rate, diminishing performance.  The NexION 300 Q ICP-MS is specified to require 100~--140~L/s of air flow, which is supported with an external exhaust fan, installed and flow-restricted by the manufacturer.  The exact flow rate is not known. The fume hood is seemingly capable of even higher flow rates and its exhaust flow is restricted such that total exhaust rate is a reasonable fraction of the the FFU flow rate, as monitored by cleanroom curtain deflection. This ensures that the cleanroom performance is not severely diminished.  Total exhaust rates are estimated to be as high as one third of the total FFU air flow rate when the fume hood and ICP-MS are operating.  Clearly significant upgrades are possible by addition of FFUs.

\section{Procedures and Analysis}
 
Three samples of copper were obtained from Aurubis AG from the production process of  electrolytic, LME grade A, NA-ESN brand copper. This material was specially selected by request with consideration for high radio-purity requirements and was controlled for reduced production of cosmogenic isotopes.    The first sample, which I will refer to as sample 1, was the anode material used as the input to Aurubis's electrolytic purification process, where anode copper is dissolved in an electrolyte, based on sulfuric acid, and deposited on a cathode using an electrical current.   Impurities are separated at both the anode and cathode depending on the electrochemical properties of the impurities.  The anode copper is the first cast of the copper after the pyrometallurgic process. Sample 2 was taken directly from the cathode of the electrolysis without further processing.  Sample 3 was  a final commercial product, rod of 8~mm diameter produced by casting and rolling the cathode material.  According to Aurubis this material was also processed in a furnace.  The precise details of the production process are not made publicly available.  Samples were obtained from multiple points in the production process with an idea that differences in contamination levels might provide information about sources of contamination. 

Sample handling, calibration and background-control procedures were established for quantitative analysis of Th and U content of the samples by ICP-MS, with sensitivity goals near 1\ppt.  Since samples are digested and thus diluted in acid, this requires sensitives of Th and U in solution on the level of $\rm 10^{-14}g/g$, near the intrinsic background limit of the NexION-300.

\subsection{Vessel preparation}

Copper was digested into solution in strong nitric acid in 60~ml PFA digestion vessels from Salivex. Contaminants adsorbed onto or diffused into the surface of the dissolution vessels can be leached out by the digestion acid, producing background signals.  Fisher Scientific Trace Metal Grade nitric acid comes with a certificate of analysis certifying levels of Th and U below $\rm 1\cdot10^{-8}g/g$, and I have observed Th contamination in this acid to be near this level.   I found that cleaning digestion vessels with this quality of acids alone was not sufficient to reduce backgrounds to below or near the intrinsic sensitivity level of the analysis.  I used direct measurements of backgrounds in acid blanks to guide development of suitable vessel cleaning procedures.  PFA vessels were found as a standard digestion vessel in the industry.  The selection of PFA is supported by evidence of low radioactive impurity levels achievable in PFA itself~\cite{Leonard08}, and, based on limited trial, I found desirable background levels to be more easily achievable using the PFA vessels compared to standard laboratory glassware.  

For this study, all sample vessels were initially new and then used only for one sample.  However, vessels have been re-cleaned since. Blank solutions in the re-cleaned vessels produce background levels of Th and U indistinguishable from vessels used for the first time.   Cleaning procedures for all new glassware and PFA digestion vessels were strictly defined in three stages as a matter of general lab practice. As with all other work reported here, all cleaning work was performed in the cleanroom while wearing cleanroom suits, boot covers, hair covers, face-masks, and powder-free gloves.  In this work trace-metal grade \hnot refers specifically to Fisher Scientific Trace-Metal grade \hnot and \uph refers to TAMAPURE-AA-100, 55\%  \hnot by mass.  Cleaning procedures are described here:

Level 1 cleaning, for new glassware:

\begin{compactenum}
\item Rinse three times inside and out in type I ultra-pure water.
\item Soak glassware in acetone in ultrasonic cleaner for at least 15 minutes. Skip this step for PFA vessels.
\item Rinse three times inside and out in type I ultra-pure water.
\item Soak in electronics grade methanol in ultrasonic cleaner for at least 15 minutes. For PFA use only a brief rinse. 
\item Rinse three times inside and out in type I ultra-pure water.
\item Soak glassware in 3 to 1 dilution of 70\% \hnot, reagent grade, for 15 minutes.
\item Rinse three times inside and out in type I ultra-pure water.
\end{compactenum}

Level 2 cleaning, suitable for many calibration and testing purposes:  

\begin{compactenum}
 \item Start with vessels cleaned with level 1 procedures.
 \item Rinse three times inside and out in type I ultra-pure water.
 \item Fill vessels with concentrated trace-metal grade HNO3. 
 \item Let vessels sit with acid overnight in the heated ultrasonic cleaner set to 70~\cels  programmed for at least 1 hour of agitation.  Preferably use a clean secondary container with Type I water and/or dilute acid for sonicating.
 \item Rinse vessel exteriors with \hnot.
  \item Rinse three times inside and out in type I ultra-pure water.
	\item Leave vessels in a covered container, ventilated while drying.
\end{compactenum}
	
Level 3 cleaning, suitable for sample digestion and analysis with best possible sensitivities:

\begin{compactenum}
\item Clean vessel with level 2 cleaning, both vessels and lids for PFA containers.
\item Rinse three times inside and out in type I ultra-pure water.
\item Set vessels in a container cleaned level 2 (or close) procedures.
\item Fill vessel with 20\% to 30\% \uph, then to top (overflowing) with type I water.
\item Screw down vessel lids, firmly, for PFA sample vessels.  
\item Fill outer container with diluted trace-metal grade acid.
\item Sonicate at 70~\cels for one hour.  Leave at 70~\cels for at least 24 hours.
\item Rinse three times inside and out in type I ultra-pure water.
\item Leave vessels in a covered container, ventilated while drying.
\end{compactenum}

Acids measured from level 2 cleaned vessels generally produced backgrounds above intrinsic levels:
Any expedited drying is performed with 99.999\% compressed Ar gas since it is readily available for use with ICP-MS. 

In practice, diffusion of contaminants into fresh acid was tested before beginning acid digestion of samples.

\subsection{Sample Digestion and Analysis}

For digestion and analysis, pieces of each copper sample were cut to sub-samples of approximately three or four grams in mass.  Before digestion for analysis, the outer surfaces of the resulting samples were removed by a strong initial acid etch. For this purpose, each sample was placed in a separate level-3 cleaned 100~ml glass beaker.  The samples were rinsed, were then covered in type I water, and \uph was added to about 15\% \hnot concentration.  The beakers were heated to approximately 60~\cels on a hotplate to accelerate etching.  Etching was performed for a few minutes until the surfaces had visibly receded, before discarding the acid and rinsing multiple times with type I water.  The water was drained, and the samples were transferred to PFA digestion vessels, rinsed, drained again and left to dry.  

The samples were then fully digested for analysis by adding concentrated \uph to the PFA vessels. The screw top lids of the PFA vessels were affixed loosely during digestion to minimize contamination, but, since the acid digestion generates gas, the lids were not tightened.  No heat was applied during digestion.

To avoid contamination, all liquids were poured directly from their stock bottles (directly from the purifier dispenser in the case of water) with no intermediate measurement vessels or pipettes. Masses were determined by measuring changes in weight of the digestion vessels on an analytic balance. As a consequence, and because the reaction occurs quickly, it can be more convenient to observe the final mass after digestion and gas loss than to attempt to to measure the added acid mass directly.  The final solution mass anyway gives the relevant information to determine the copper concentration in the solution.

Digestion of copper in nitric acid can occur by several reactions depending on the acid concentration and temperature~\cite{InorgChem}.  In practice the digestion probably occurs by a changing combination of reactions as the acid is consumed or the temperature fluctuates.  Because of uncertainty in the reaction chain, and differences in mass of gasses lost in each reaction, it was not feasible to precisely determine the minimal amount of acid for digestion of the copper.  In order ensure that enough acid was added to fully digest the copper, I assumed that the reaction goes by the following equation: 

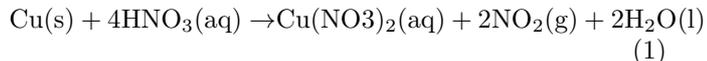
\begin{equation}
{\rm Cu(s) + 4HNO_3(aq) \rightarrow } {\rm  Cu(NO3)_2(aq) + 2NO_2(g) + 2H_2O(l) }
\label{eq:fast}
\end{equation}

which, of all the possible reactions, requires the most \hnot per gram of copper and, after accounting for gas losses, also results in the highest required net mass increase by addition of sufficient acid.  This implies that 7.22~g of \uph solution at 0.55\% concentration is needed to dissolve one gram of acid, or that a 5.77 gram net mass increase is required, per gram of copper, after gas loss.

For sample 1, the un-purified anode sample, following digestion with about 15\% more acid that nominally required, I observed small amounts of a dark substance precipitating on the vessel bottom after a couple of days.  Partially for this reason, digestion of samples for final analysis was done with significantly more acid than was required, by two or three times, although it was not clear that this suppressed precipitation in sample 1.  The precipitate was likely related to  large concentrations of several contaminants which were found by ICP-MS mass scans. The purified samples, samples 2 and 3, did not produce visible precipitates.

After digestion, the acid solutions were diluted to bring dissolved solid concentrations down to levels acceptable for analysis. I discuss this in detail below.  The dilution was performed by discarding part of the digestion solution and adding type I water, producing about 50~g of solution for analysis. While it was possible to precisely determine the sample dilution, because of chemistry ambiguities mentioned above, it was not possible to precisely know the acid content of the diluted solutions. I estimate that, for the measurements reported here, there was between between 6\% to 8\% ,by mass, un-reacted \hnot in the diluted solutions.  

A spectator blank was prepared in a fourth PFA vessel, also pre-cleaned with level 3 cleaning.  The blank was prepared simply by allowing concentrated acid to sit in the vessel in similar conditions to the samples, followed by a similar dilution step.

Quantitative analysis of the samples for concentrations of Th and U required sensitivity calibration and background subtraction.  In principle backgrounds can come from many sources including desorption in the vessel or the memory effects in the ICP-MS (desorption in the introduction system).  These could be well monitored using the spectator process blank. As well, the apparatus itself produces backgrounds which do not appear to be related the existence of any actual contaminants of the masses of interest.  These backgrounds are flat as a function of mass, with no peaks.  They do increase when a sample is introduced into the plasma, and could be related to de-focused ions of the wrong masses.  I will refer to these as intrinsic backgrounds.

The intrinsic backgrounds, as well as the sensitivities to Th and U, can change very significantly as the condition of the apparatus changes or its tuning parameters (gas flow rate, deflector voltage, etc.),  are changed.  Furthermore at high sample concentrations the sensitivities depend very strongly on the makeup of the sample being analyzed.  For this reason the signals obtained from the process blank cannot be subtracted directly from the sample signals nor can a separate calibration sample be relied upon to quantify sensitivity to Th and U in the copper samples.  Calibration was performed by adding a small amount of calibration solution to the sample solutions, minimally affecting the overall composition of the sample solutions.  This can be done either by splitting the samples, or by adding the calibration as a final step.  This is a risk vs resource decision as cleaned vessels for splitting the samples require costly acids and time to prepare.  For this work the calibration solution was added at the end of analysis, although in some cases a split of the sample had already been prepared on the side. The process blank was treated as an entirely separate measurement and was also calibrated by addition of an internal standard.

Specifically calibrations with roughly $\rm 5 \cdot 10^{-11}~g/g$ concentrations of Th and U were prepared by dilution of  
of stock  PerkinElmer Smart Tune Solution which contained $\rm 1.00~\mu g/L$ of each of Th and U (as well other elements) and .05\%~\hnot by mass.  Care was taken to ensure that the solution was acidified to an \hnot concentration of at least $\rm 0.1~mol/L$ at all times to avoid adsorption of the calibration elements onto the glass.  Thus, all dilutions were performed by adding \hnot before adding ultra-pure water.  The calibration solutions were prepared in level 3 cleaned 100~ml glass beakers and were prepared on the day of data taking.  I have found these diluted solutions to remain stable within 3\% statistical uncertainty, relative to the stock solutions, for at least several days.  Calibrations prepared in this way were found to produce linear spectrometer responses, within statistical uncertainties ($~5$\% at the lowest concentration), from $2.5\cdot10^{-13}~g/g$ to $1\cdot10^{-9}~g/g$. 

Before performing the final analysis, test analyses were made using cathode copper to investigate the maximum beneficial concentration of copper within the analysis solution.  I compared two samples, one with  1.4\% dissolved copper by mass and a second with 6.4\% copper.  Sensitivities to Th and U in the solution were about three times lower for the 6.4\% solution than the 1.4\% solution.  Relative to the copper mass this represents a modest factor of 1.5 improvement in sensitivity for the higher concentration.  However, as more copper is injected into the plasma it builds up on the torch glass and torch cones, and the stability and sensitivity of the spectrometer both degrade continuously, but not always steadily.  At higher concentrations this occurs more quickly.  This limits available analysis time, thus directly reducing overall statistical power and/or the number of samples that can be analyzed.  The instability can increase systematic uncertainty greatly, and analysis generally becomes very difficult.  Once the sensitivity becomes unacceptable or the spectrometer becomes too unstable, the sample introduction system, torch glass, and cones must be disassembled, cleaned, re-assembled, and the spectrometer must be re-tuned. The spectrometer injection system was cleaned in this way after these test analyses but prior to final analysis. 

The apparatus was re-tuned (settings were optimized) using the above mentioned Smart Tune Solution.  Use of stock smart tune solution at $\rm 1\cdot10^{-9}~g/g$ concentrations results in Th and U emanating from the injection system, producing background signals.  Restoring backgrounds to intrinsic levels required flushing the injection plumbing overnight with approximately 20\%~\hnot.  Most of this flushing was performed with the plasma off. Even at levels of $\rm 5\cdot10^{-11}~g/g$ a couple of hours of flushing can be required, so calibrations used on the day of analysis were limited to even lower concentrations.  In general it is probably preferable and certainly possible to perform initial device optimization with solutions diluted to about $\rm 5\cdot10^{-11}~g/g$.

Specifically I attempted to optimize deflector voltage, plasma RF power, plasma gas flow,  and auxiliary gas flow.  A combination of manual and automated optimizations were used with feedback from the Th and U sensitivities.  Customized method files (these define the mass scan, scan timing, etc.) were developed to show these signals in real time for manual optimization, and the default method files for automated tuning were modified to feedback specifically on the \Th{232} and/or \U{238} signals. Several nearby masses were also monitored, such as mass (m/z) 236. and half-integer masses such as 237.5, in order to monitor background levels while tuning.  This procedure produced sensitivities to \Th{232} and \U{238} equivalent to about $\rm 50,000~counts/s$ for calibration samples of $\rm 1\cdot10^{-9}~g/g$ concentrations.  Throughout the course of sample analysis, several hours, the RF power was re-optimized to compensate for degradation in performance related to sampling of solutions with high concentrations of copper.

Along with the primary sample solutions, five auxiliary solutions were prepared and on-hand for final analysis:
\begin{compactenum}  
\item Calibration sample prepared with ${\rm 5.0\cdot10^{-11}~g/g}$ of Th and U. 
\item Level-2 cleaned quartz beaker containing a fraction of the Aurubis sample-3 analysis solution, with  ${\rm 4.4\cdot10^{-11}~g/g}$ concentrations of Th and U.  
\item Level-3 cleaned class beaker containing type I ultra-pure water.
\item  Level-3 cleaned PFA container with \uph at roughly 15\% \hnot concentration, for system rinsing.	
\item  Level-3 cleaned PFA container with \uph at roughly 15\% \hnot concentration, for a reference blank.
\item The sample process blank as previously described. 
\end{compactenum}

Vessel~1 was used to prepare spiked solutions in lower concentrations but was never itself sampled for analysis during the day. Vessel~2 was used to monitor the ICP-MS sensitivity to a realistic sample, perform initial daily tuning, and to make minor tuning corrections, but not for quantitative analysis.

The samples were taken up into the ICP-MS nebulizer through a thin plastic tubing by the built-in peristaltic pump operating at 20~rpm (measured to be $\rm 13.2~cm^3/hr$).  A typical analysis sequence started with analysis of vessel 2 to check and or optimize sensitivity.  For a spiked solution such as this, I would then rinse the sampling tube briefly in vessel~3 and then flush the ICP-MS sampling system using vessel~4, usually monitoring backgrounds with live data acquisition.  After reducing \Th{232} and \U{238} backgrounds to levels at or near intrinsic levels, I would take background data while sampling from vessel~5.  The progression of blanks reduces cross contamination before ultimately recording process blank data from vessel~6. 

After obtaining satisfactory sensitivities and tunes, a copper sample solution was analyzed. Because these all proved to be very low in Th and U, consistent or marginally consistent with zero, a brief rinse and flush of the sampling system was sufficient before switching to a different sample.  Since the sensitivity of the apparatus changed throughout the course of the day, generally getting slowly lower, this cycle was repeated a few times to help monitor those sensitivities.  

The ICP-MS was operated in quantitative analysis mode, outputting uncalibrated pulse counts. All solutions were treated in the software as standard foreground samples, with calibration and background subtraction performed by offline data analysis.  All masses were monitored in peak hopping mode, meaning data was taken only at the precise mass specified, without scanning the mass peak.  This produces data at the peak centroids, thus maximizing data rate.  Several masses were scanned near the masses of interest, sweeping over all specified masses usually 30 to 60 times per reading, depending on the desired statistics or reading time. Masses (m/z) 232 and 238 were scanned for foreground signals of Th and U respectively.  Masses 231.5 and 232.5 were scanned as a measure of intrinsic background of the 232 signal, and masses 235.5 and 238.5 were scanned for intrinsic background subtraction of the mass 238 signal.  Mass 237.5 was not used because of the presence of a small peak at mass 237.  Mass 237 is consistent with PbNO which seems reasonable given the existence of high levels of lead and nitric acid.  All of these peaks were scanned with dwell times of 500~ms per sweep each, yielding 30 seconds of total integration time per mass for a 60-sweep reading.  Typically 4 readings were averaged for each data point reported here, representing two minutes of integration time per mass. 

Mass 208 (Pb) was also scanned, usually with a much lower dwell time of 5~ms.  All copper digestion samples contained lead contamination (much higher levels in the anode sample) and this signal proved to be extremely useful for monitoring the sensitivity and stability of the instrument in real-time during sample measurements and in some cases for rejecting data if extreme changes were noticed.  For arbitrary samples, intentional introduction of a tracer element could be very useful for this purpose.

\section{Results and Conclusions}

Data for all samples and the process blank are shown in table~\ref{tab:data}. By using spectral background subtraction from measurements at half-integer masses, the confidence level for a signal detection is determined purely by the combined counting statistics of all data used, without influence from systematic measurement-to-measurement sensitivity deviations.  Fluctuations in sensitivities, specifically fluctuations between samples and spiked calibrations, then result only in an overall calibration scale error.  For sample 2, the first sample measured during the day, sensitives varied less than 2\% throughout analysis of the sample, before and after the calibration spike.  This behavior has been reproducible in separate analysis sessions, so ideal results can be achieved for critical samples by limiting analysis to one sample.  For samples 1 and 3, fluctuations in the lead sensitivities were observed.  These fluctuations were used to make quantitative corrections, resulting conservatively, in both cases, in  increased limits.  There is a clear correlation between Pb sensitivities and Th and U sensitivities. However, as this correlation is not firmly established, a systematic scaling error equal to the size of the corrections was also applied to these data and is indicated in table~\ref{tab:data}.  A minimum but conservative 10\% scaling error is applied to account for calibration uncertainties arising from comparison of new and old standard solutions.  

\begin{table*}[pbt!]
\begin{tabular}{	r|		p{5cm}|				c			|			c			|			c			|			c		|	}
\hline																												
		&		&						&						&					&						\tabularnewline
		&		&	Sample 1					&	Sample 2				&	Sample 3					&	Process Blank					\tabularnewline
		&		&	Anode Copper					&	Cathode Copper				&	"Cast \& rolled copper"				&	(acid +water)					\tabularnewline
\hline																												
	1	&	sample mass, pre-etch	[g]&	2.6929					&	not recorded				&	4.1497					&						\tabularnewline
	2	&	sample mass, post etch	[g]&	2.2078					&	2.9014					&	3.8217					&						\tabularnewline
	3	&	dissolved copper concentration	&	1.08\%					&	1.23\%					&	1.12\%					&	assume 1\%					\tabularnewline
	4	&	Spike, ppt of Th and U in solution	&	2.34					&	2.74					&	3.38					&	1.49					\tabularnewline
	5	&	Th sensitivity [c/s/ppt] \newline 
(dilution corrected/ roughly lead rate corrected) 	&	0.111		$\pm$		0.007	&	0.072		$\pm$		0.002	&	0.033		$\pm$		0.001	&	0.152		$\pm$		0.006	\tabularnewline
	6	& U sensitivity [c/s/ppt] \newline
(dilution corrected/ roughly lead rate corrected) 	&	0.122		$\pm$		0.008	&	0.091		$\pm$		0.002	&	0.041		$\pm$		0.001	&	0.165		$\pm$		0.007	\tabularnewline
	7	&	Th signal [ppt of sample]	&	11.6		$\pm$		1.1	&	2.2		$\pm$		0.5	&	4.0		$\pm$		1	&	7.1		$\pm$		0.7	\tabularnewline
	8	&	U signal [ppt of sample] 	&	11		$\pm$		1.0	&	1.5		$\pm$		0.4	&	1.4		$\pm$		0.6	&	5.7		$\pm$		0.6	\tabularnewline
	9	&	intrinsic Th bkgd [ppt equivalent]\newline 
(231.5 + 232.5)	&	8.1		$\pm$		0.6	&	1.9		$\pm$		0.3	&	1.9		$\pm$		0.06	&	6.7		$\pm$		0.5	\tabularnewline
	10	&	intrinsic U bkgd [ppt equivalent] \newline
(235.5 + 238.5)	&	7.7		$\pm$		0.6	&	1.4		$\pm$		0.3	&	3.6		$\pm$		0.7	&	5.8		$\pm$		0.4	\tabularnewline
	11	&	Th concentration  [ppt]	&	3.2		$\pm$		1.3	&	0.23		$\pm$		0.6	&	2		$\pm$		1.3	&	0.22		$\pm$		0.88	\tabularnewline
	12	&	U concentration  [ppt]	&	3.2		$\pm$		1.2	&	0.047		$\pm$		0.49	&	-2.3		$\pm$		1	&	-0.15		$\pm$		0.77	\tabularnewline
	13	&	Th limit [ppt]	&		$<$4.0				&		$<$ 1.2				&		$<$ 4.1				&		$<$ 1.6				\tabularnewline
	14	&	U limit  [ppt]	&		$<$ 5.2				&		$<$ 0.9				&		$<$ 1.7				&		$<$ 1.3				\tabularnewline
	15	&	Systematic scale error	&		30\%				&		10\%				&		25\%				&		30\%				\tabularnewline
\hline
\end{tabular}
\caption{Data from ICP-MS analysis of Aurubis, LME grade A, NA-ESN brand electrolytic copper production samples and the process blank.  Sample mass is shown before and after pre-digestion surface etch. Final copper concentration in the analysis solution is indicated.  Sensitivities are shown for \Th{232} and \U{238} as measured in the analysis solutions, by spiking the solution with Th and U to the levels indicated in ppt (parts-per-trillion, $10^{-12}~g/g$ ).  The \Th{232} and \U{238} signals are shown in units of ppt of copper (not ppt of solution). Background levels measured at half interval masses are shown in the same units. The final results show the background subtracted signals.  Uncertainties shown are $\rm 1~\sigma$ from Gaussian counting statistics alone. Limits are one-sided 95\%CL, simply estimated as the mean plus $\rm 1.64~\sigma$. Systematic scaling uncertainties are estimated in row~15 and are primarily due to instabilities observed in the lead signals. }
\label{tab:data}
\end{table*}

For comparison, portions of the same three copper samples were also sent for analysis by the generally less sensitive technique of glow discharge mass spectroscopy (GDMS) at the National Research Council of Canada (NRC).  Limits of $<2.0\cdot10^{-8}~g/g$ of Th and U were reported for all three samples, consistent with the ICP-MS results of this work.  Scans of other elements also showed qualitative consistency, although the ICP-MS scans for other elements were not quantitatively calibrated. 

The limits reported in table~\ref{tab:data} for Th and U in Aurubis NA-ESN copper are useful verification of the radiopurity of this material, and thus the suitability of this copper for use in low-background rare-event searches in particle physics.  Specifically the low limits set on Th and U in the anode, the un-purified copper, hint at the possibility that the final consumer product may have contamination levels significantly below $\rm 1\cdot10^{-12}g/g$.  Although not systematically studied, other contaminants such as Pb were observed to be reduced in concentration by multiple orders of magnitude in the purified samples relative to the anode sample.

The results reported here demonstrate the ability to set measurement limits of Th and U on the order of $\rm 1\cdot10^{-12}g/g$ using direct analysis of acid digested samples.  The technique relied on the use of high solid concentrations, high acid concentrations for flushing sampling plumbing, use of spectral-based background subtraction, sensitivity monitoring from matrix contaminants, and calibrated internal calibrations.  Work was performed in a cost-effective cleanroom environment of simple construction.

In comparison to techniques~\cite{grinberg} involving chemical concentration processes, some systematic concerns such as blank subtraction and chemical recovery efficiency are removed.  The disadvantage of this simplified procedure is that sample throughput rate is limited by the need to clean the sample introduction system frequently.  For dedicated single-sample analysis the technique can produce limits dominated by counting statistics with $1~\sigma$ uncertainties significantly below $\rm 1\cdot10^{-12}g/g$.  This method might not be suitable for high-throughput labs offering routine analysis services.  However, it provides a useful alternative for achieving sensitivities to low concentrations of Th and U in a research-oriented environment, specifically for application to the development of low-background detectors for nuclear and particle physics.







\begin{acknowledgments}
I thank Andreas Piepke at the University of Alabama for coordination with Aurubis and CNRC.  I am grateful to Dr. J Schmidt and Dr. F. Meyer at Aurubis for patience and help with discussion and selection of copper samples. I thank Patricia Grinberg at CNRC-INMS in Canada, and Balz Kamber and Jacques Farine at Laurentian University for their valuable consultation related to ICP-MS analysis. This work was supported by the 2013 Sabbatical Year Research Grant of the University of Seoul and by the NRF.

\end{acknowledgments}


\end{document}